\documentclass[12pt, draftcls, onecolumn, a4paper]{IEEEtran}

\usepackage{fancybox, fancyhdr}
\usepackage{xurl}
\usepackage{hyperref}
\hypersetup{colorlinks,allcolors=blue}
\usepackage{enumerate, enumitem} % Numbered Bullets
\usepackage{environ}
\usepackage{cite}
\usepackage{amsmath,amssymb,amsfonts, multirow, eurosym}
\usepackage{algorithmic}
\usepackage{graphicx}
\usepackage{textcomp, booktabs}
\usepackage[english]{babel} % English Formatting
\usepackage[utf8]{inputenc} % Standard Encoding
\usepackage{multicol} % Multi Column Environment
\usepackage{flafter} % Reference any 'float'
\usepackage{soul} %to highlight text
\usepackage{color} %to highlight text
\usepackage{array} %to highlight text
\usepackage{tikz} %to highlight text

\usepackage{caption}
\usepackage{subcaption}

\newcommand\copyrighttext{%
  \footnotesize \textcopyright 2022 IEEE. This paper is accepted for publications in IEEE Transactions on Industrial Informatics. Personal use of this material is permitted.
  Permission from IEEE must be obtained for all other uses, in any current or future
  media, including reprinting/republishing this material for advertising or promotional
  purposes, creating new collective works, for resale or redistribution to servers or
  lists, or reuse of any copyrighted component of this work in other works.
  DOI:10.1109/TII.2022.3182988}
\newcommand\copyrightnotice{%
\begin{tikzpicture}[remember picture,overlay]
\node[anchor=south,yshift=10pt] at (current page.south) {\fbox{\parbox{\dimexpr\textwidth-\fboxsep-\fboxrule\relax}{\copyrighttext}}};
\end{tikzpicture}%
}

\begin{document}

\title{A Functional Architecture for 6G \\Special Purpose Industrial IoT Networks}

\author{Nurul Huda Mahmood, Gilberto Berardinelli, Emil J. Khatib, Ramin Hashemi, Carlos de Lima, and~Matti Latva-aho% <-this % stops a space
\thanks{N. H. Mahmmod, R. Hashemi and M. Latva-aho are with 6G Flagship, Centre for Wireless Communications, University of Oulu, Finland.}
\thanks{G. Berardinelli is with the Department of Electronic Systems, Aalborg University, Denmark.}
\thanks{E. J. Khatib is with the University of Malaga, Spain.}
\thanks{C. de Lima is with Nokia Mobile Networks, Oulu, Finland.}
\thanks{This research has been supported by the Academy of Finland, 6G Flagship program under Grant 346208}
\thanks{Corresponding e-mail: \href{mailto:nurulhuda.mahmood@oulu.fi}{nurulhuda.mahmood@oulu.fi}.}
%\thanks{Manuscript received xx; revised xx}
}

%\markboth{IEEE Transactions on Industrial Informatics}{Mahmood \MakeLowercase{\textit{et al.}}: A Functional Architecture for 6G Wireless IIoT Networks}
%\markboth{IEEE Transactions on Industrial Informatics}%
%{A Functional Architecture for 6G Wireless IIoT Networks}
%{Mahmood \MakeLowercase{\textit{et al.}}: 6G Wireless IIoT Network Functional Architecture}



\maketitle
\thispagestyle{empty}
\copyrightnotice

\begin{abstract}
Future industrial applications will encompass compelling new use cases requiring stringent performance guarantees over multiple key performance indicators (KPI) such as reliability, dependability, latency, time synchronization, security, etc. Achieving such stringent and diverse service requirements necessitates the design of a \textit{special-purpose Industrial Internet of Things (IIoT) network} comprising a multitude of specialized functionalities and technological enablers. This article proposes an innovative architecture for such a special-purpose 6G IIoT network incorporating seven functional building blocks categorized into: \textit{special-purpose functionalities} and \textit{enabling technologies}. The former consists of \textit{Wireless Environment Control}, \textit{Traffic/Channel Prediction}, \textit{Proactive Resource Management} and \textit{End-to-End Optimization} functions; whereas the latter includes \textit{Synchronization and Coordination}, \textit{Machine Learning and Artificial Intelligence Algorithms}, and \textit{Auxiliary Functions}. The proposed architecture aims at providing a resource-efficient and holistic solution for the complex and dynamically challenging requirements imposed by future 6G industrial use cases. Selected test scenarios are provided and assessed to illustrate cross-functional collaboration and demonstrate the applicability of the proposed architecture in a wireless IIoT network.
\end{abstract}


% \fi
%\IEEEpeerreviewmaketitle

%%%%%%%%%%%%%%%%%%%%%%%%%%%%%%%
%%%%%%%%%%%%%%%%%%%%%%%%%%%%%%%
%%%                         %%%
%%%         SECTION         %%%
%%%                         %%%
%%%%%%%%%%%%%%%%%%%%%%%%%%%%%%%
%%%%%%%%%%%%%%%%%%%%%%%%%%%%%%%

\section{Introduction}
\label{sec:intro}

\IEEEPARstart{W}{ireless} network evolution follows a trend of having a new generation every decade. Following this trend, the fifth generation (5G) New Radio (NR) introduced in the 2020's is expected to be succeeded by the sixth generation (6G) wireless network around 2030. The International Telecommunications Union (ITU), which is responsible for defining International Mobile Telecommunications (IMT) systems, has already started to examine future technology trends for ``IMT towards 2030 and beyond''\footnote{\href{https://www.itu.int/en/myitu/News/2021/02/02/09/20/Beyond-5G-IMT-2020-update-new-Recommendation}{Beyond 5G: What's next for IMT?}}; with the first set of definitions expected to be available around mid-2024. The standardization body 3GPP also plans to initiate studies into 6G requirements from mid-2024 with the first basic 6G standard anticipated to be defined around 2027. 

Alongside conventional human type communications, 5G NR has incorporated two dedicated service classes to support machine type communications (MTC), namely massive MTC and ultra-reliable low-latency communications (URLLC). MTC will play a dominant role in future beyond 5G/6G systems owing to its huge potential for business and technological innovations. From a business perspective, the ability to automate communications between machines paves the way towards \textit{connectivity as a service} business model, thereby enabling a wide range of novel applications and use cases~\cite{MAS+22_factory5G, MTCwhitePaper2020}. New technological innovations are also urgently needed to allow intelligent, scalable and energy efficient solutions that can meet the challenging requirements of future MTC networks~\cite{MBA+22_IIoT}. 

Industrial Internet of Things (IIoT) primarily caters to MTC applications with requirements from an industrial network such as a manufacturing setup and control systems for railways and energy management~\cite{MBA+22_IIoT, MMG+21_IIoT}. IIoT converges information and communication technology with operational technology, and is an integral component of the fourth industrial revolution (Industry 4.0), where cyber-physical systems utilize reliable and fast control loops between sensors and actuators to automate delicate control tasks~\cite{berardinelli2018_wirt}. Communications networks designed to provide seamless connectivity for various IIoT applications are realized using a wide range of, mostly proprietary, wired and wireless solutions~\cite{MAS+22_factory5G}. In many cases, such solutions are customized to the local use case motivated by security considerations, technology limitations and legacy design, which limit their scalability and adaptability to different situations~\cite{berardinelli2018_wirt}. The flexibility and mass customization of production envisioned by Industry 4.0 requires agile and open connectivity solutions while maintaining the high performance guarantees accorded by existing customized solutions~\cite{CLH+21_multirobot}. The introduction of URLLC service class in 5G NR is the first step towards having an universal wireless standard to meet these needs. 

Although the 5G NR URLLC solutions have made progress in enabling wireless connectivity for IIoT use cases, the true vision of replacing proprietary connectivity solutions with a universal wireless system in industrial networks is yet to be realized~\cite{adeogun2020towards}. Moreover, emerging use cases in the coming decade will impose new requirements that were not considered in 5G NR. A robust, scalable, and efficient 6G network is thus necessary to meet the diverse requirements of the upcoming decade. Research efforts towards defining 6G have been pursued by academic researchers~\cite{6gWP2019, DAS_19_whatShould6G, MTCwhitePaper2020, MBM+21_mtcEurasip}, major international research projects~\cite{URB+21_hexaX_6G} and key industrial players~\cite{viswanath6G_2020, samsung6g}. Due to the diversity of use cases and application areas, 6G is expected to be a `network of networks' that will aggregate multiple types of resources connecting at different scales~\cite{URB+21_hexaX_6G}. 

This paper contributes to the ongoing efforts in defining 6G by proposing an integrated functional architecture for a 6G special-purpose IIoT network. In particular, we focus on wireless IIoT networks motivated by the demand for robust wireless communication technologies that can seamlessly replace wired connections in the IIoT domain~\cite{MAS+22_factory5G}. A transmission failure in an IIoT network may lead to downtime, which is costly in terms of money and efforts. Hence, there is an enormous value - not only in terms of academic research, but also from a commercial perspective - in designing robust IIoT networks. 

We propose an architecture consisting of seven specialized functions grouped into \textit{special-purpose functionalities} and key \textit{enabling technologies}. Such an architecture will allow optimizing the implementation of the key functionalities through dedicated network components. We envision that the proposed 6G special-purpose IIoT network will be a part of the wider 6G ecosystem of \textit{Network of networks}. The foundation for such networks dedicated to specific use cases like IIoT is rooted in the \textit{Non-public (i.e. private) networks (NPN)} feature in the 3GPP Release-16 standard~\cite{GMB19:5Gevolution, MBA+22_IIoT}, and the URLLC/IIoT enhancement features to be finalized in the forthcoming Release 17. Although some of the \textit{special-purpose functionalities} and \textit{enabling technologies} presented in this article have already been discussed in several 6G related articles, to the best of our knowledge, there is no existing literature discussing a framework integrating all these functionalities into a system concept and highlighting the tight integration among them. Furthermore, a few of the functionalities presented in this work, such as \textit{high-accuracy positioning through integrated communications and sensing}, are rather novel concepts that have started being discussed in the literature just recently~\cite{LCM+22_ISAC}. The novel contributions of this work with respect to the state-of-the-art can be summarized as follows:
\begin{itemize}
    \item We take a \textit{clean slate} approach and present a functional architecture for 6G special-purpose IIoT networks that can efficiently meet the stringent and diverse design goals. We believe that the distribution of different enabling technology components across different functionalities within the proposed architecture make its management and integration agile and seamless. 
    \item The tight integration among the presented functionalities enabling them to meet the stringent requirements of the future industrial communications networks is detailed. We limit ourselves to a selected set of features relevant for IIoT applications since designing a complete system architecture requires multi-disciplinary expertise and collaboration that is beyond the scope of this work.
    \item We demonstrate the feasibility of the proposed functional architecture through numerical evaluations, thus moving beyond merely presenting the concepts in generic terms.
\end{itemize}

The rest of the paper begins with an overview of the key requirements of a 6G IIoT special purpose network in Section~\ref{sec:scenario}. Each of the functional blocks in the proposed architecture is then elaborated in Section~\ref{sec:functional}. The effectiveness of the proposed architecture in enhancing the network performance is demonstrated with an example use case in Section~\ref{sec:results}, describing how the different blocks interact for the sake of controlling the radio environment to improve IIoT connectivity. Finally Section~\ref{sec:conc} concludes the paper.

%%%%%%%%%%%%%%%%%%%%%%%%%%%%%%%
%%%%%%%%%%%%%%%%%%%%%%%%%%%%%%%
%%%                         %%%
%%%         SECTION         %%%
%%%                         %%%
%%%%%%%%%%%%%%%%%%%%%%%%%%%%%%%
%%%%%%%%%%%%%%%%%%%%%%%%%%%%%%%

\section{Key Requirements of 6G IIoT Networks}
\label{sec:scenario}

The notion of multi-service communications introduced in 5G is expected to expand further in 6G with the emergence of new use cases and service classes driven by advances in communications and other technologies such as sensing, imaging and artificial intelligence (AI)~\cite{samsung6g, MTCwhitePaper2020}. The potential use cases and expected KPIs of future 6G networks have recently been considered by various authors (e.g.,~\cite{MTCwhitePaper2020, 6gWP2019, viswanath6G_2020, DAS_19_whatShould6G}). The diverse set of expected 6G use cases can be described in terms of a set of KPIs representing \textit{i)} ultra-reliability, \textit{ii)} low-latency, \textit{iii)} high data-rate, \textit{iv)} massive access, \textit{v)} energy-efficiency and \textit{vi)} localization and sensing accuracy. However, all of these -- sometimes conflicting -- requirements are not needed to be met simultaneously in a single system. Ultra-reliability, low-latency, localization and sensing accuracy, and to a certain extent high data-rate, will be the most important KPIs for future 6G IIoT networks, which is the focus application in this work. This section briefly summarizes some use cases and requirements relevant to future 6G IIoT networks. 

The key requirements and expected KPIs of a 6G IIoT special purpose network stem from its various potential use cases. For instance, the application of digital twins (DT), which provide a real-time representation of physical objects in the virtual world, will be further extended to include digital representation of the (wireless propagation) environment and assets beyond manufacturing - leading to the \textit{massive DT} use case \cite{URB+21_hexaX_6G}. This will require a new class of URLLC with high data rates, thereby imposing novel design challenges. As another example, wirelessly connected multirobot systems in smart factories with highly reliable driving systems (i.e. navigation, collision avoidance, sensing, etc.) will increase the demand on the scale, complexity and QoS of the connectivity requirement~\cite{khatib2021optimization, CLH+21_multirobot}. 6G will bring further advances in private networks \cite{ahokangas2021platform, MBA+22_IIoT}, where the ownership of the network infrastructure and data belongs to the entity that is using it to support its own applications. Special-purpose networks often go hand in hand with private networks. IIoT is a typical example where special-purpose networks are used, since it is often a good practice, both in technical and managerial terms to separate the critical business resources from the rest. 

The diverse set of 6G KPIs will either evolve from the main 5G KPIs or are novel ones introduced in 6G. Important 5G KPIs like reliability and latency will be further extended in 6G to consider the broader concept of \textit{dependability}, whose main attributes are availability, reliability, safety, integrity, and security~\cite{NHS+21_dependability, MAM+22_missionEffCap}. Similarly, emerging applications like connectivity for multirobot systems including unmanned aerial objects will render connectivity requirements to be measured in volumetric unit. In addition, 6G will witness a number of novel KPIs coming to the fore. Many emerging applications requiring precise localization will render localization accuracy an important KPI~\cite{LCM+22_ISAC}. Alongside, sophisticated industrial control applications will require tight synchronization, low jitter and time-sensitive communications~\cite{GRN21_tsn, MBA+22_IIoT}. A summary of the prospective IIoT relevant KPIs in 6G and their comparison with 5G KPIs is presented in Table~\ref{tab:KPIs}. 

%------------------------%
%        Table           %
%------------------------%
\begin{table}[t!]
\begin{center}
\caption{A comparison of selected MTC KPIs in 5G and 6G}
\label{tab:KPIs}
\begin{tabular}{p{3.6cm} c c }
\toprule
KPI & 5G target & 6G target \\
\midrule
Per radio link reliability~\cite{hexaXd1_3} 	& $1 - 10^{-5}$ 	& $1 - 10^{-9}$ \\
Application level E2E reliability~\cite{MAS+22_factory5G} 	& \textit{not considered} 	& $1 - 10^{-6}$ \\
Per radio link latency~\cite{hexaXd1_3} 		& $1$ ms 		& $0.1$ ms \\
Application level E2E latency~\cite{MAS+22_factory5G} & $5$ ms	& $<1$ ms \\
Connection set-up time & \textit{not considered} 	& $<1$ ms \\
Connection density~\cite{MTCwhitePaper2020} & $1$ device/m$^2$ & up to $10$ device/m$^3$ \\
Spectral efficiency (downlink) & $\sim25$ bpcu & $\sim40$ bpcu \\
Device lifetime~\cite{MTCwhitePaper2020} & $10$ years & $40$ years \\
Energy consumption~\cite{MTCwhitePaper2020} & \textit{low} & \textit{ultra-low} \\
Positioning accuracy~\cite{LCM+22_ISAC} & $30$ cm & $1$ cm/$5$ mm \\
Jitter~\cite{5GACIA_integration5G_tsn} & $1 \mu$s & $<0.1 \mu$s \\
E2E optimization~\cite{CLH+21_multirobot, MAS+22_factory5G} & \textit{not considered} & \textit{relevant} \\
Dependability~\cite{hexaXd1_3} & \textit{not considered} & \textit{relevant} \\
\bottomrule
\end{tabular}
\end{center}
%\vspace{-6mm}
\end{table}

%%%%%%%%%%%%%%%%%%%%%%%%%%%%%%%
%%%%%%%%%%%%%%%%%%%%%%%%%%%%%%%
%%%                         %%%
%%%         SECTION         %%%
%%%                         %%%
%%%%%%%%%%%%%%%%%%%%%%%%%%%%%%%
%%%%%%%%%%%%%%%%%%%%%%%%%%%%%%%

\section{Key Functional Blocks}
\label{sec:functional}

\begin{figure*}[t]
    \centering
    \includegraphics[width=0.8\linewidth]{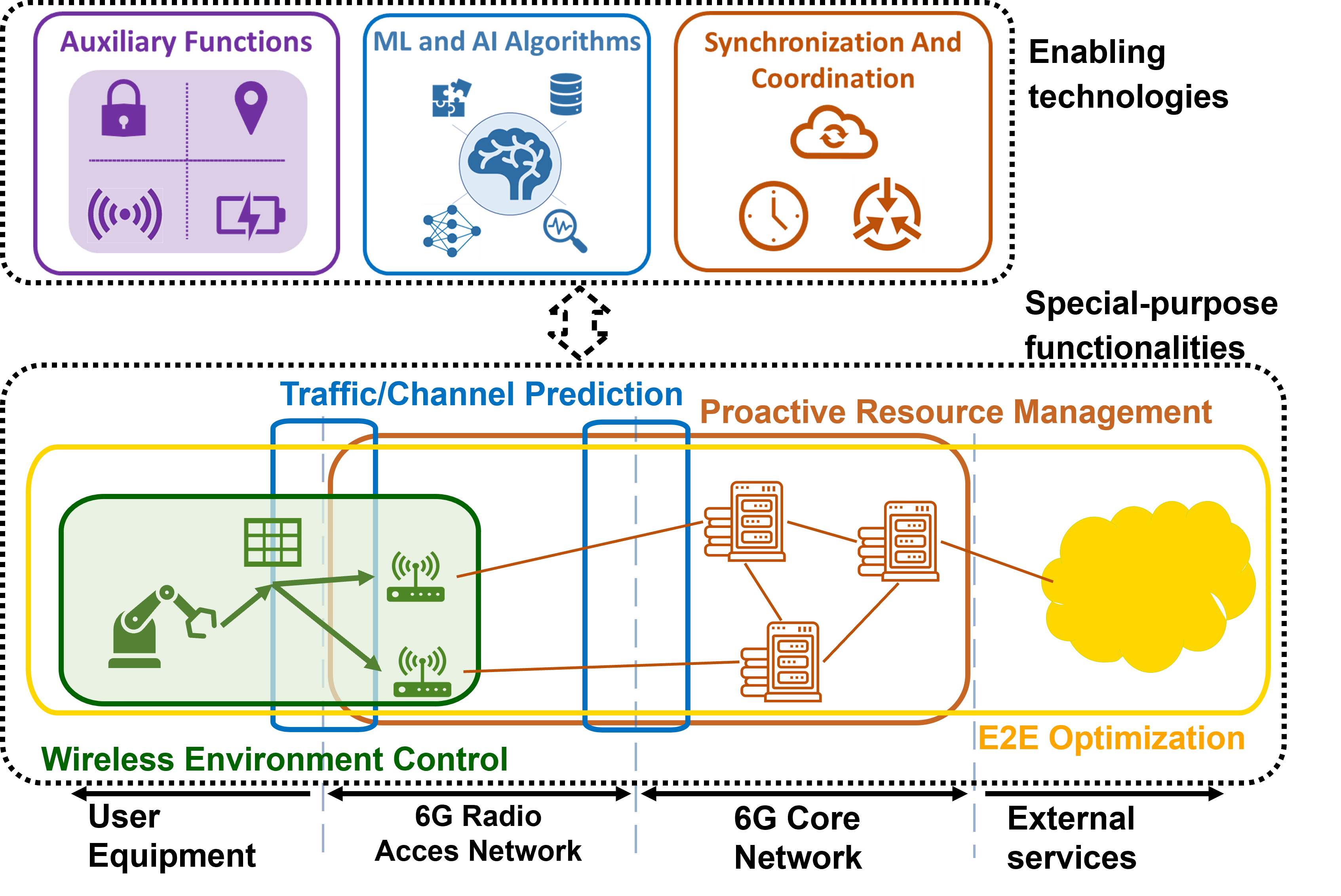}
    \caption{The proposed functional architecture for a 6G special purpose IIoT network}
    \label{fig:architecture}
\end{figure*}

This section presents the proposed functional architecture for a 6G IIoT network, comprising of seven specialized functional blocks, as illustrated in Fig.~\ref{fig:architecture}. The 6G network is divided into four sections (user equipment (UE), radio access network (RAN), core network (CN) and external services). The proposed architecture is categorized into \textit{special-purpose functionalities} and \textit{enabling technologies}. The former includes \textit{Wireless Environment Control}, \textit{Traffic/Channel Prediction}, \textit{Proactive Resource Management} and \textit{End-to-End (E2E) Optimization}. The \textit{Wireless Environment Control} functionality deals mainly with the air interface, that is, the part of the UE and RAN directly related to radio communications. The \textit{Traffic/Channel Prediction} functionality analyzes the traffic occurring both in the air interface and internal interfaces within the network, which is closely related to the services provided over the network. The \textit{Proactive Resource Management} functionality orchestrates the resources in the 6G network, such as the scarce radio resources and computing nodes in the RAN and the CN. The \textit{E2E Optimization} functionality monitors and improves all the elements in all the sections of the network in order to provide service-specific optimizations and increase the E2E QoS. On the other hand, the \textit{special purpose functionalities} rely on a set of enabling technologies that provide the computational tools (\textit{Machine Learning (ML) and AI Algorithms}) and services (\textit{Auxiliary Functions}) that compose their operation, as well as a \textit{Synchronization and Coordination} functionality that orchestrates their work, overviews their effects and prevents conflicts. The proposed architecture focuses on the lower layers of the OSI protocol layer, and higher layer network management functionalities are beyond the scope of this work.

The proposed special-purpose 6G IIoT network is envisioned to be a private network deployed in a industrial setting. For instance, it can be a completely private network deployed, owned and operated by the factory owner; a private network deployed and (partially) operated by a network operator, network equipment vendor or a third party at the premise of, and in collaboration with, the factory owner; or a virtual private network operating on a sliced portion of a public commercial cellular network~\cite{MAS+22_factory5G}. Such a special-purpose 6G IIoT network will be supported by the envisaged spectrum usage scenarios for future 6G networks, where spectrum may be allocated to a confined area in line with the \textit{Network of networks} concept~\cite{ahokangas2021platform}. 

%------------------------%
%      sub section       %
%------------------------%
\subsection*{\bf Special Purpose Functionalities}

\subsection{Wireless Environment Control}
\label{sub:environment}

\subsubsection*{Functionality definition}
The stochastic nature of the wireless environment and random interference are among the key challenges when ensuring high reliability in wireless communications. We envision a dedicated functionality addressing these challenges. Generally, fluctuation of the received signal power due to fast fading is tackled through diversity techniques, such as multiconnectivity~\cite{mahmood2019resource}. However, not all forms of diversity solutions are applicable in industrial scenarios. In order to tame the random fading in wireless channels, novel approaches that can track, predict and even favorably shape the propagation condition are anticipated towards 6G.

\subsubsection*{Research directions and enabling techniques}
Modeling the wireless environment through a DT and/or with the assistance radio frequency (RF) (e.g., ray-tracing), non-RF technologies (e.g., color and depth images) and integrated sensing and communications techniques can be used to better track and predict the channel behavior~\cite{LCM+22_ISAC}. An accurate wireless map can provide deterministic knowledge of the path loss and also allow better prediction of the channel fading~\cite{ZX21_channelKnowledgeMapping}. Similarly, preceding RF signals and depth-images can be exploited for received power prediction in wireless channels, so that resulting outages due to deep fades or channel blockages can be predicted beforehand and managed proactively. Recently, reconfigurable intelligent surface (RIS) aided wireless communication is gaining widespread attention as a means to engineer the propagation environment~\cite{DZD+20_RIS}. An RIS is made of meta-materials with adjustable phase shift and amplitude response. By adjusting these, an RIS can passively transform the random wireless environment into a controllable channel, thereby rendering the signal propagation more deterministic. Additionally, centralized information about the activities of different transmitters can lead to better information about the interference generation and its variation across time and space. Advanced ML and AI algorithms can be applied to reliably track the interference variation, and predict its future values. Thus, guarantees on the interference experienced by different receivers can be provided with high accuracy, which leads to better performance guarantees and efficiency~\cite{MLA+20_predictive}. 

\subsubsection*{Interaction with other functionalities}
This functional block closely interacts with and aids the implementation of \textit{Proactive Resource Management} and \textit{E2E Optimization} functionalities, whereas itself is enabled by the \textit{Auxiliary Functions} (e.g., localization) and, to a certain extend, \textit{ML and AI Algorithms}.

%------------------------%
%      sub section       %
%------------------------%

\subsection{Traffic/Channel Prediction}
\label{sub:traffic}

\subsubsection*{Functionality definition}
IIoT devices vary widely in their functionalities, leading to diverse traffic types in the industrial network. The traffic generated by a given node can be temporally and/or spatially correlated. A detailed modeling of such spatio-temporal traffic correlation, and possibly collating it with other contextual information, can be utilized to better predict the arrival of the source traffic. This in turn leads to more efficient access to the shared wireless media, better resource utilization, lower latency and higher reliability. 

\subsubsection*{Research challenges and enabling techniques}
Traffic in IIoT networks can be broadly categorized as periodic and event-triggered. Predicting periodic traffic arrival requires identifying the exact time instance when the traffic will be generated, and its size. There is a rich body of literature investigating mathematical models for such traffic behavior, such as the non-homogeneous Poisson process model and Markovian processes and frequency based analysis~\cite{wang2015approach} for long-term correlation modeling. However, such linear prediction models are not well suited to predict event-driven IIoT traffic. ML-based prediction methods have recently emerged as viable solutions to capture the complex and non-linear spatio-temporal dependence structure of the event-driven traffic generation. The list of the proposed solutions include deep belief network based prediction method, applying a long short-term memory structure~\cite{HM20_lstm}, and applying deep convolutional neural networks to model the spatial and temporal traffic dependence~\cite{ZZY+18_cnn}. Despite their demonstrated performance under specific scenarios, most ML based solutions are very much scenario-specific, dependent on large training data, and not easily amenable to interpretation. To address this, the main operational goal for the proposed prediction functionality is to integrate learning based prediction with accurate modeling of the spatial-temporal traffic correlation considering the available domain knowledge and contextual information. Such an approach in 6G will likely lead to more accurate and explainable prediction models in 6G.

\subsubsection*{Interaction with other functionalities}
We envision this functional block to be receiving data from multiple UEs, RAN elements such as access points, and the CN interfaces, while interacting with the \textit{Auxiliary Functions} and the \textit{ML and AI Algorithms} as the key enablers. The traffic patterns predicted by this functionality will serve as inputs to the \textit{Proactive Resource Management} and \textit{E2E Optimization} functionalities.

%------------------------%
%      sub section       %
%------------------------%

\subsection{Proactive Resource Management}
\label{sub:RRM}
\subsubsection*{Functionality definition}
The preceding two blocks discuss methods to predict the traffic arrival, and track the channel and interference variation. The {\textit{Proactive Resource Management} block} explores tools and techniques to utilize the predicted traffic and interference information in designing scheduling and interference management techniques. The main goals are to reduce the access and transmission latency through prompt scheduling, ensure high reliability through intelligent interference management schemes and guarantee efficient utilization of scarce resources. Towards this end, network intelligence technologies running at different elements across the network that fully automate the network operations will become the standard in 6G~\cite{BGF+22_networkIntelligence, MBA+22_IIoT}. 

\subsubsection*{Research challenges and enabling techniques}
Conventional reactive resource management principles, where traffic is served on-demand and transmission failures are addressed through retransmissions, are not well suited to meet the stringent latency and reliability requirements of many IIoT use cases. Instead, intelligent mechanisms that proactively allocate resources by predicting the traffic demand and the channel conditions beforehand are needed~\cite{MBA+22_IIoT}. Scheduling and assignment of transmission resources, e.g., transmit power, bandwidth, etc., are the two main aspects of resource management at the physical and the access layers. A mismatch between the traffic arrival and the allocated slots for existing proactive scheduling techniques like semi-persistent scheduling leads to under-utilized resources or additional access latency. Such concerns can be addressed by dynamically grouping users based on their predicted traffic profile and allocating resources accordingly. For example, users with low traffic correlation may access the same channel using grant-free schemes, leading to significantly reduced collision probability.

Predictive resource allocation is more challenging than scheduling, especially in the case of an interferred network. Predicting the received signal and interference strength can result in a more accurate prediction of the future signal-to-interference plus noise ratio (SINR), which in turn leads to improved link adaptation~\cite{MLA+20_predictive}. In the case of centralized resource allocation or coordination between base stations (BS), the predicted strongly interfering BS can be proactively requested to manage the interference through muting or silencing a specific resource block. The proactive resource allocation paradigm extends beyond the lower layers to include proactive allocation of network slicing and edge computing resources. Resource allocation for network slicing is particularly challenging since the physical network and edge computing resources can be particularly scarce, and due to the limitations imposed by the mobility requirements of the end-user~\cite{BVS+20_slicingRA, BGF+22_networkIntelligence}. In this context, proactive resource allocation by forecasting the end-user demand and dynamically matching it with the available resources leads to better QoS, and fairer and more efficient allocation of the scarce resources~\cite{FP20_NSmeetsPT, MBA+22_IIoT}.

\subsubsection*{Interaction with other functionalities}
The \textit{Proactive Resource Management} functionality logically lies at the center of the proposed functional architecture. On one hand, it is driven by system requirements including E2E optimization constraints and the network traffic profiles. On the other hand, it relies upon the \textit{Synchronization and Coordination}, \textit{Wireless Environment Control} and \textit{Auxiliary Functions} to implement the outputs of its ML and AI powered algorithms. 

%------------------------%
%      sub section       %
%------------------------%

\subsection{End-to-end Optimization}
\label{sub:E2E}
\subsubsection*{Functionality definition}
A cellular network is a very complex system with each deployed hardware and software element having numerous configuration parameters, whose optimization is a multidimensional problem. Once an optimal configuration point is achieved, it must be actively maintained, since the environmental conditions of the network change over time and some elements may malfunction. Optimization involves iteratively adjusting the parameters of interest until reaching a desired performance. This scheme can target functionalities at different levels, for instance, at the RAN, the CN or the radio interface level.

\subsubsection*{Research challenges and enabling techniques}
The latest network optimization trends include an E2E perspective, where the monitored performance indicators cover specific data sessions or application classes, and represent the QoS at the application layer. At the same time, the optimized parameters affect all the network elements that serve the corresponding application~\cite{KFV+21_E2Eco}. E2E optimization also gains relevance with functions such as Multi-access Edge Computing (MEC)~\cite{abbas2017mobile}, where network computing resources are moved ``closer'' to the end users to reduce the response time of certain services. With the help of ML/AI functions running on Big Data resources of the block described in Section \ref{sub:ML_AI}, these functions can be implemented, replicated for diverse applications and reused for novel optimization loops.

Current E2E optimization functions cover the communication channels between the endpoints of a data session. Nevertheless, the problems and sub-optimal behavior may come from the configuration of the end devices. The detection of such issues can be very important to discard non-existing network problems that may reflect in certain performance indicators (such as the data throughput). In the future, standardized interfaces may even allow the network to propose configuration improvements to the endpoints, such as buffer sizes or priority queue parameters. The full-scale E2E connectivity landscape will also include integration of wired and wireless networks to support the multitude of challenging QoS requirements~\cite{MAS+22_factory5G}. 

\subsubsection*{Interaction with other functionalities}
Since the E2E communication link spans across the entire network, the \textit{E2E Optimization} functionality has to interact with all other functionalities, including the enabling technologies.

%------------------------%
%      sub section       %
%------------------------%

\subsection*{\bf Enabling Technologies}

%------------------------%
%      sub section       %
%------------------------%

\subsection{Synchronization and Coordination}
\label{sub:sync}

Traditionally, wireless communication systems distribute time reference among base stations using the global navigation satellite system (GNSS) infrastructure which is enough to meet present-day time requirements \cite{art:godor-icsm20}. 5G NR implements two main procedures to propagate time reference across communicating devices, namely the typical GNSS-based approach and over-the-air synchronization. 5G NR synchronization requirements are not necessarily more restrictive, though high-accuracy time synchronization has become a critical aspect owing to the frame structure and advanced radio resource algorithms. GNSS/global positioning system time synchronization provides a reasonable time reference, though the performance deteriorates in canyon or indoor deployment scenarios. In this regard, transport network over-the-air synchronization becomes a promising alternative. Moreover, 5G NR coordinated RAN features such as coordinated multi point transmission, beamspace processing and time-of-flight-based positioning demand much tighter time synchronization.

In order to provide reliable data transport, as well as seamless connectivity of sensors, actuators and controllers, special purpose IIoT networks demand accurate time and clock synchronization so as to harmonize time among often independent and distributed clocks. For example, programmable logical controllers are typically employed in sequential processes and thus need very tight clock synchronization with time variation limited to at most $1~\mu$s. Legacy industrial settings are typically deployed through wired networks implementing time-sensitive networking (TSN) standards~\cite{MAS+22_factory5G}. Hence, the proper integration of 5G and TSN requires E2E time synchronization to support time-critical industrial applications, which can be implemented via generic precision time protocol~\cite{art:godor-icsm20}.
%------------------------%
%      sub section       %
%------------------------%

\subsection{ML and AI Algorithms}
\label{sub:ML_AI}

The \textit{ML and AI Algorithms} functional block perform actions to solve a problem based on information extracted from an environment that is either too complex or has too many variables to be efficiently solved by conventional optimization algorithms. This is precisely the case for many problems in the 6G network, where management functionality is bound to have both complex and large datasets given the great amount of users, service transactions and infrastructure components~\cite{MBA+22_IIoT}. In ML algorithms, the operation can be separated in a learning and an exploitation phase. The learning phase is usually slow, requires large datasets and is computationally costly, whereas the exploitation, that is, using the model to obtain new outputs based on unobserved inputs, is usually based on simple and fast operations. For the learning phase there are three main possible locations: distributed (i.e., running in the terminals, using only the dataset visible to the computing node), federated (i.e., the algorithm runs in the devices and no data is shared among the nodes, but models are shared, improving the information available at each node) or centralized (data is sent to the cloud, where the ML algorithms run). For the exploitation, that is, the use of the learned model to obtain new estimations, the algorithm can either run in the device, where it will produce fast results at relatively low computational cost, or in the 6G core network or network edge~\cite{CLH+21_multirobot}.

%\subsubsection*{Enabling techniques}
The objective of this block is to simplify the development and deployment of novel ML-based applications. Centralizing the learning phase within the 6G network will have several advantages; firstly, the management of learning datasets is easier and enables the creation of novel functions quickly. Furthermore, it allows access to a larger dataset compared to a decentralized implementation. Secondly, the possibility of creating libraries and reusing code increases the productivity of future development. Finally, centralization allows for more efficient usage of computing resources, including Big Data technologies \cite{khatib2016self}, such as Cloud Computing, schema-free databases, etc. The exploitation phase can be done, as earlier indicated, in the network edge or even in the terminals, so response times of ML-based services are not compromised~\cite{BGF+22_networkIntelligence}.

As time has progressed, ML has been adopted in a wider set of applications in many diverse fields. This has led to the emergence of a need for trustworthiness. While classical ML algorithms act as a black box, that take datasets as input and produce a model as an output, there is a growing demand for explainable ML \cite{rai2020explainable}, i.e., that the produced model is somehow justified. There are two approaches for achieving this; either using glass box (or white box) ML algorithms, that is, algorithms that are self-explanatory, or using algorithms that, attached to a black box ML algorithm can obtain a human-understandable explanation. The development and use of these systems in 6G (and, therefore, in the present block) are key for a widespread adoption, based on the trust of operators and end users.

\subsection{Auxiliary Functions}
\label{sub:auxilliary}

5G NR offers very promising features such as large bandwidth, very high carrier frequency, massive antenna arrays and densification. B5G/6G systems are expected to continue this trend and further develop them. This will be enabled by advanced RAN infrastructure and auxiliary network functionalities beyond communication technologies. Future wireless networks will thus increasingly rely on auxiliary side-information for its overall operation and performance optimization. These include, high-accuracy localization and high-resolution environment mapping through integrated sensing and communications to support adaptive resource allocation~\cite{LCM+22_ISAC}, dynamic/intelligent spectrum management, as well as security enablers at the lower layers of the protocol stack.

As depicted in the functional architecture in Fig.~\ref{fig:architecture}, this block will gather relevant information and then implement such auxiliary functions. In this context, new ML/AI solutions have the potential to leverage the availability of sensing information to streamline operational procedures and optimize the overall performance. Thus, aiming to support very high reliability and extreme low latency, real-time and accurate sensing becomes a crucial component. In this aspect, the research challenges are related to tradeoffs between sensing and communications, dual-functional (sensing and communications) waveform design, and signal processing algorithms to address mutual interference cancellation~\cite{LCM+22_ISAC}. Similarly, ML/AI-enabled intelligent spectrum management has the potential to not only cope with spectrum scarcity and harsh interference, but also to significantly improve spectrum allocation, especially in the context of NPNs and licensed shared access~\cite{MAS+22_factory5G}. In such deployments, collaborative- and edge-computing will become increasingly indispensable thus raising concern about the proper implementation of privacy, security and trust. In particular, the network need to be made more resilient to security vulnerabilities at the CN-edge-RAN communication links, privacy issues in private networks (especially with shared network infrastructure), and advanced attacks employing ML/AI techniques~\cite{mika_wp10, MBA+22_IIoT}.

%------------------------%
%      sub section       %
%------------------------%

\subsection*{\bf Deployment Scenarios and Implementation Costs}
\label{sub:cost}

One of the main advantages of the proposed functional architecture is that it is modular, i.e., each block can be implemented independently while still being interconnected with the other blocks. Hence, the proposed functionalities can be placed centrally or distributed across the network. The blocks that require more prompt actions, such as \textit{wireless environment control}, can be located in close proximity to the end devices, e.g. on a edge cloud. On the other hand, functionalities operating over a more relaxed time scale can eventually run in a central cloud, therefore benefiting from higher computational capacity, besides offloading the edge cloud. As many of the functionalities may rely on ML and AI algorithms, the models can eventually be trained in the central cloud, and then transferred to the edge cloud for their prompt execution. The models can be then further optimized online by relying on persistent connections between the edge and the central cloud. By offloading complex computing tasks and data storage from the devices to the edge and/or the cloud, edge computing further enhances the operation of limited terminals not only as a way to reduce latency, but also to save energy and carry out highly demanding computational procedures. There are, of course, several tradeoffs in exchange of this advantage:
\begin{itemize}
    \item The proposed architecture will not be readily applicable in other scenarios, e.g., massive IoT networks. Different configurations and functionalities of 6G will have to be adapted to those uses.
    \item There is an added communication overhead, especially when implemented as a distributed architecture. This overhead affects both the devices (that will have more functionalities available, but also will need to rely more on the network connectivity), and the network elements (edge computing nodes, remote servers, etc.).
    \item The computational cost for the network operator will be higher and the operator has to dedicate more resources to automatic functions that allow to deploying, maintaining and optimizing a much more complex network than prior generations.
\end{itemize}

Nevertheless, these tradeoffs allow devices to have more functionalities available from the network, and efficient development of new services. Hence, the computational and communication costs of the proposed architecture can be very efficiently managed, while reaping its advantages. 

%%%%%%%%%%%%%%%%%%%%%%%%%%%%%%%
%%%%%%%%%%%%%%%%%%%%%%%%%%%%%%%
%%%                         %%%
%%%         SECTION         %%%
%%%                         %%%
%%%%%%%%%%%%%%%%%%%%%%%%%%%%%%%
%%%%%%%%%%%%%%%%%%%%%%%%%%%%%%%

\section{Example of Usage of Proposed Architecture}
\label{sec:results}

Novel technologies like the use of RIS are introducing a paradigm shift by allowing to shape the propagation characteristics favorably, and hence are poised to be an integral part of future 6G systems~\cite{DZD+20_RIS, HAM+22_jointRIS_ICL}. An RIS-assisted wireless link allows the components of the received signal at a target receiver to add constructively resulting in boosting the received signal-to-noise ratio (SNR). In this section, we illustrate how the proposed architecture works for the usage and optimization of phase shifts of an RIS to improve the connectivity quality of an IIoT network. We consider an industrial scenario where a BS serves one or more actuators. Due to the presence of significant noise and interference caused by large machinery and heavy multipath propagation effects induced by highly reflective structures, the wireless channel in an industrial environment is much different compared to conventional wireless propagation scenarios~\cite{Che16_industrialChannel}. 

\begin{figure}[ht]
  \centering
  % include first image
  \includegraphics[width=.6\columnwidth]{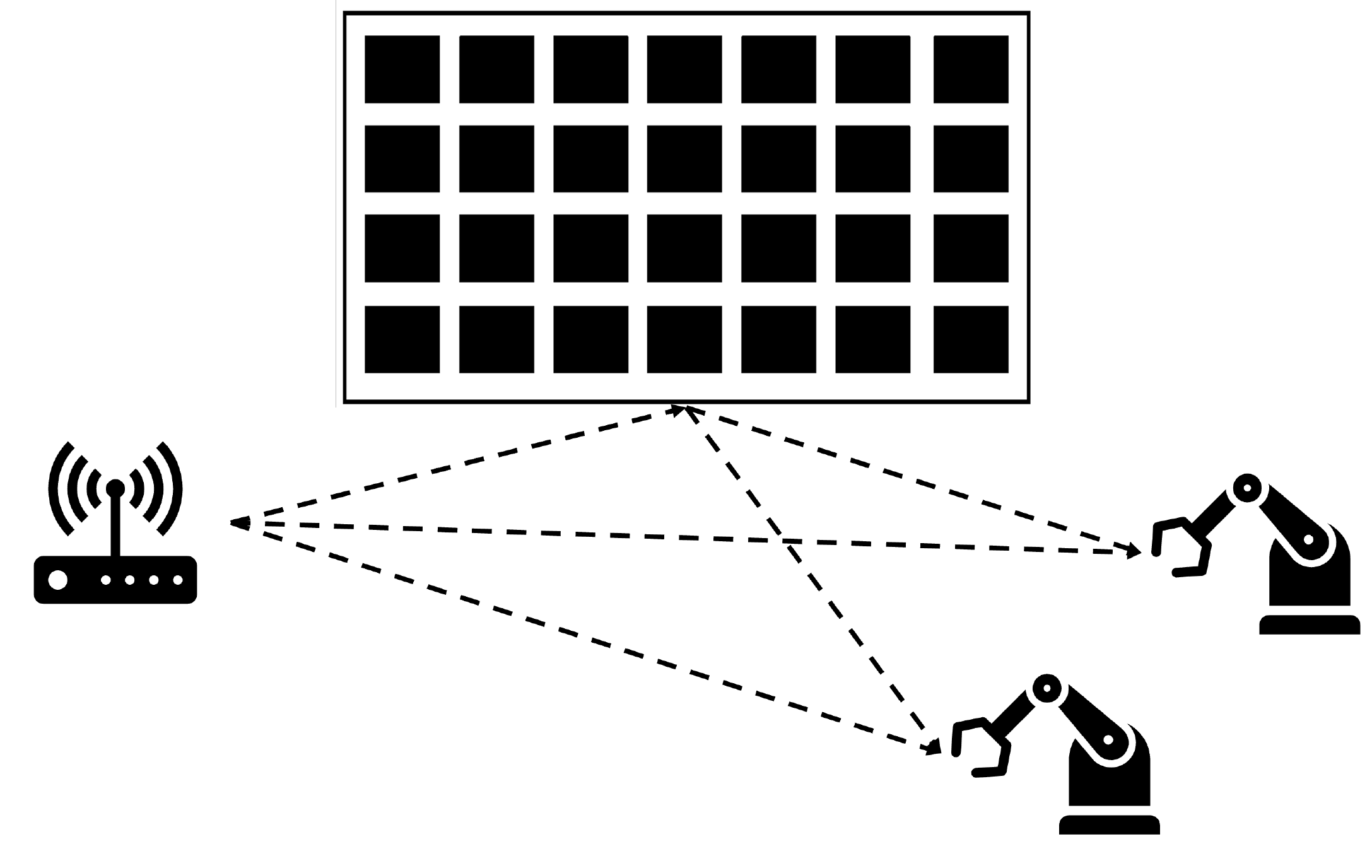}  
  \caption{The considered system model where a BS serves multiple actuators in a wireless industrial environment.}
  \label{fig:risSysModel}
\end{figure}

We investigate the use of RIS to control the wireless environment with the objective of mitigating the adverse propagation effects considering the deployment scenario presented in Fig.~\ref{fig:risSysModel}. RIS represents an instance of the \textit{Wireless Environment Control} functionality. In this example, we show the interconnection of \textit{Wireless Environment Control} block with the \textit{Traffic/Channel Prediction} block, that has the scope here of predicting channel state information (CSI) as this is fundamentally important for leveraging RIS, as well as with the \textit{ML and AI Algorithms} block, that has the task of optimizing the phase response of the RIS. 

\subsection{Application of the \textit{Wireless Environment Control} Block}
\label{sub:idealRIS}

In the first illustrative example, we demonstrate how an RIS can be used to control the wireless environment. We assume an ideal scenario where the BS, which controls the RIS elements, has full CSI of the BS-RIS and the RIS-actuator channels. The BS optimizes the phase shifts of each of the $N$ RIS elements such that the phases of complex base-band received signals coming from the BS-RIS and the RIS-actuator channels add constructively at the receiver end. We further consider the possibility of optimizing the reflection phases without any discretization error, i.e., the phases of the RIS elements are able to take any value in the continuous range $(0, 2\pi]$ and the absence of any loss due to absorption or coupling between the RIS elements~\cite{BDD+19_RIS}. Such strong assumptions allow assessing the full potential of RIS and the corresponding results serve as performance upper bounds.

In the case of single-antenna BS and actuator, the received narrowband signal at the actuator in the presence of an RIS with $N$ elements at time instant $t$ is given by
\begin{equation}
    y(t) = \left[f(t) + \mathbf{g}(t)^H \mathbf{\Theta}(t) \mathbf{h}(t) \right]s(t) + n(t),
    \label{eq:RIS_rxSignal}
\end{equation}
where $f(t)$, $\mathbf{g}(t) \in \mathbb{C}^{N\times 1}$ and $\mathbf{h}(t) \in \mathbb{C}^{N\times 1}$ represent the combined effects of the small scale fading and the distance dependent path loss of the BS-actuator, RIS-actuator and BS-RIS channels, respectively. The $N-$dimensional square matrix $\mathbf{\Theta}(t) = \text{diag}\left(\beta_1 e^{i\theta_1}, \beta_2 e^{i\theta_2}, \ldots, \beta_N e^{i\theta_N} \right)$ represents the RIS elements with $\beta_n \in [0, 1]$ being the amplitude attenuation and $\theta_n \in (0, 2\pi]$ being the phase shifts, $\forall n \in \{1, 2,\ldots, N\}$. 

\begin{figure}[t]
    \centering
    \includegraphics[width=0.65\linewidth]{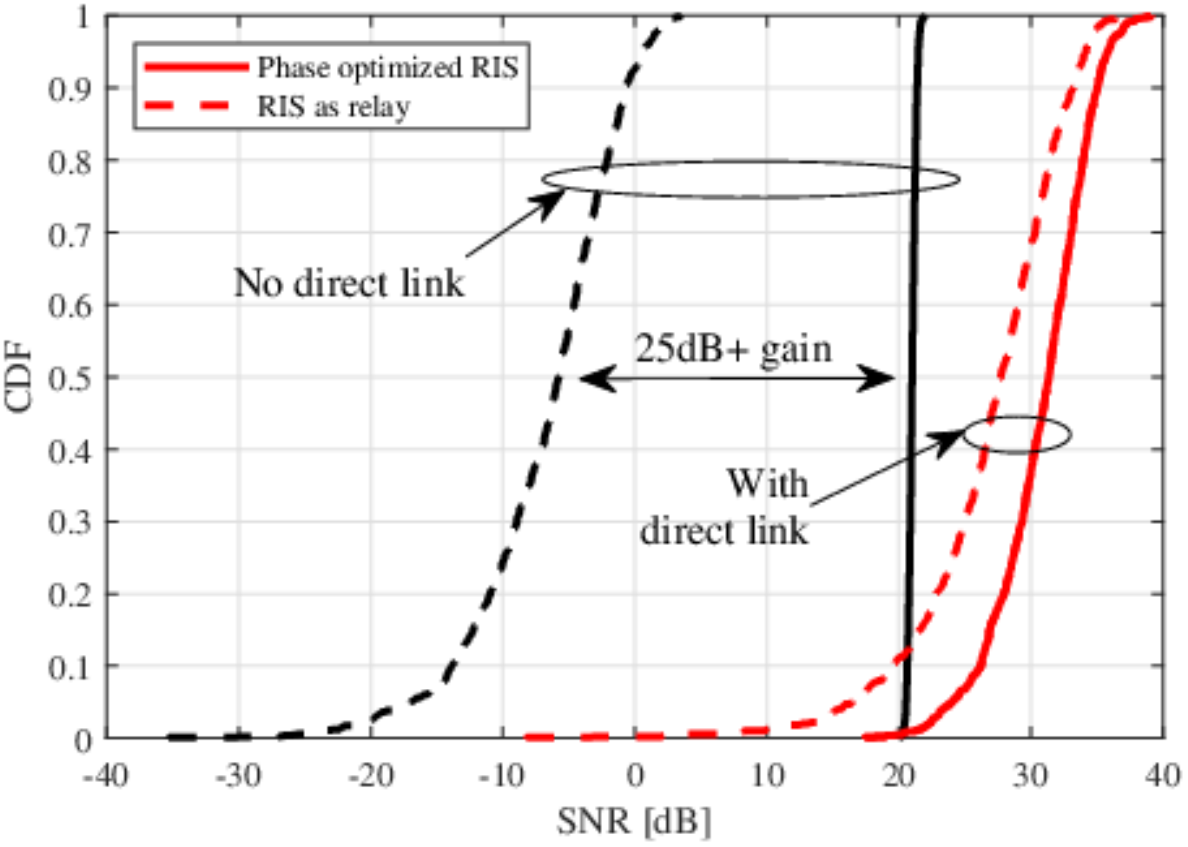}
    \caption{The SNR CDF for the phase optimized RIS as well as when the RIS acts as a relay, with and without the direct link $f(t), \, N = 512$.}
    \label{fig:RIS_snrDist}
\end{figure}

Fig.~\ref{fig:RIS_snrDist} presents the cumulative distribution function (CDF) of received SNR at the actuator in the presence of an RIS. The number of elements in the RIS, $N = 512$ and the path loss in dB is given by $\text{PL} = 34.53 + 38\log_{10}(d),$ where $d$ is the transmitter-receiver separation distance in metres. We consider a scenario where the BS, the RIS, and the actuator are located at points $(0,0), (10, 10)$ and $(100,0)$ on the $(x,y)$ grid (in meters), respectively. Four different cases are presented in Fig.~\ref{fig:RIS_snrDist}. The SNR distribution when the RIS merely acts as a relay with static phase shifts ($\theta_n = 0 \, \forall n$) considering the absence ($f(t) = 0$) and presence ($f(t) \neq 0$) of the direct link are shown by the dashed lines. The solid lines represent the SNR distribution through an RIS with optimum phase shift, $\theta_n^\ast = -(\xi_n + \zeta_n) \, \forall n$, where $\xi_n$ and $\zeta_n$ are the phase shifts of the $n^{th}$ element of $\mathbf{g}(t)$ and $\mathbf{h}(t)$, respectively.

The median and the range of the SNR, in dB, for the four different cases presented in Fig.~\ref{fig:RIS_snrDist} are tabulated in Table~\ref{tab:snrValues}. By comparing the values corresponding to the cases without the direct link (black lines), we observe that optimizing the RIS elements' phase shifts results in a significant improvement in the SNR, with a gain of over $25$ dB in the median value. Furthermore, the variation in the SNR is greatly reduced, as indicated by the steep slope of the CDF plot and the range values in Table~\ref{tab:snrValues}. The lesser is the variation in the received SNR through a given channel, the more deterministic is that channel - which implies higher dependability. Hence, the use of RIS in controlling the wireless environment has important positive implications for industrial scenarios requiring dependable communications. The gains from employing RIS are less pronounced in the presence of the direct source-destination link with a modest gain of less than $4$ dB in the median value. This is because the compound source-RIS-destination channel is reported to decay with the product of the source-RIS and RIS-destination distances, which leads to a much higher path loss compared to the direct channel~\cite{BDD+19_RIS}. Thus, the SNR gain accorded by the RIS is masked by the significantly stronger source-destination path. Nonetheless, the range of the SNR distribution is still significantly reduced by around $20$ dB, thus making the channel more deterministic.

%------------------------%
%        Table           %
%------------------------%
\begin{table}[t!]
\begin{center}
\caption{The median and the range of the SNR, in dB, for the four different cases presented in Fig.~\ref{fig:RIS_snrDist}}
\label{tab:snrValues}
\begin{tabular}{r | c c | c c }
\toprule
 & \multicolumn{2}{c|}{Phase optimized RIS} & \multicolumn{2}{c}{RIS as relay} \\
 & median & range & median & range \\
\midrule
w/ direct link & $31.35$ & $21.6$ & $27.71$ & $41.3$\\
w/o direct link & $21$ & $1.8$ & $-5.79$ & $38.97$\\
\bottomrule
\end{tabular}
\end{center}
%\vspace{-6mm}
\end{table}

\subsection{Application of the \textit{Traffic/Channel Prediction} Block}
The results presented in Section~\ref{sub:idealRIS} consider an ideal scenario and give an indication of the general performance trends. In order to evaluate the benefits of controlling the wireless environment using RIS in realistic scenarios, we evaluate the performance under non-ideal assumptions in this section. The perfect CSI assumption is relaxed. In practice, an RIS contains a large number of elements which increases the number of links whose CSI are to be estimated, resulting in an unacceptably large pilot overhead. In addition, the RIS itself is a passive component and hence the channel can only be sensed at a receiver by sounding the channel from a transmitter. This implies the need to estimate the the channels $\mathbf{g}$ and $\mathbf{h}$ in~\eqref{eq:RIS_rxSignal} from the observation of the cascaded effect of $\mathbf{g}^H \mathbf{\Theta} \mathbf{h}$~\cite{WSD21_risChannel}.

The overhead and complexity of channel estimation in an RIS-assisted wireless systems can be reduced by implementing specific algorithms in the \textit{Traffic/Channel Prediction} block. The unique properties of the RIS channel can be exploited for this purpose. For instance, given the fixed locations of the BS and RIS, the BS-RIS channel $\mathbf{g}$ usually remains unchanged over a longer period, and hence can be estimated less often compared to the RIS-actuator channel $\mathbf{h}$~\cite{WSD21_risChannel}. In the case of multiple actuators, the fact that the BS-RIS channel is the same for all users results in the cascaded channels associated with different users having some correlation among them, which can be exploited to reduce the estimation overhead. Similarly, the channel through the RIS can be transformed and tested for sparsity in the angular domain, which may arise in scenarios with limited propagation paths. In the case of such sparsity, the channel estimation problem can be formulated as a sparse signal recovery problem, and subsequently solved using the rich body of knowledge available on solving compressed sensing problems~\cite{MA21_chEstRIS}. The application of ML based approaches to directly observe the impact of the RIS elements phase shifts on the achievable rate (or other performance metrics) without estimating the channel itself has also been proposed as an alternative solution~\cite{HAM+21_RIS_URLLC}, which is further investigated in Section~\ref{sub:MLapplication}.

\begin{figure}[ht]
    \centering
    \includegraphics[width=0.65\linewidth]{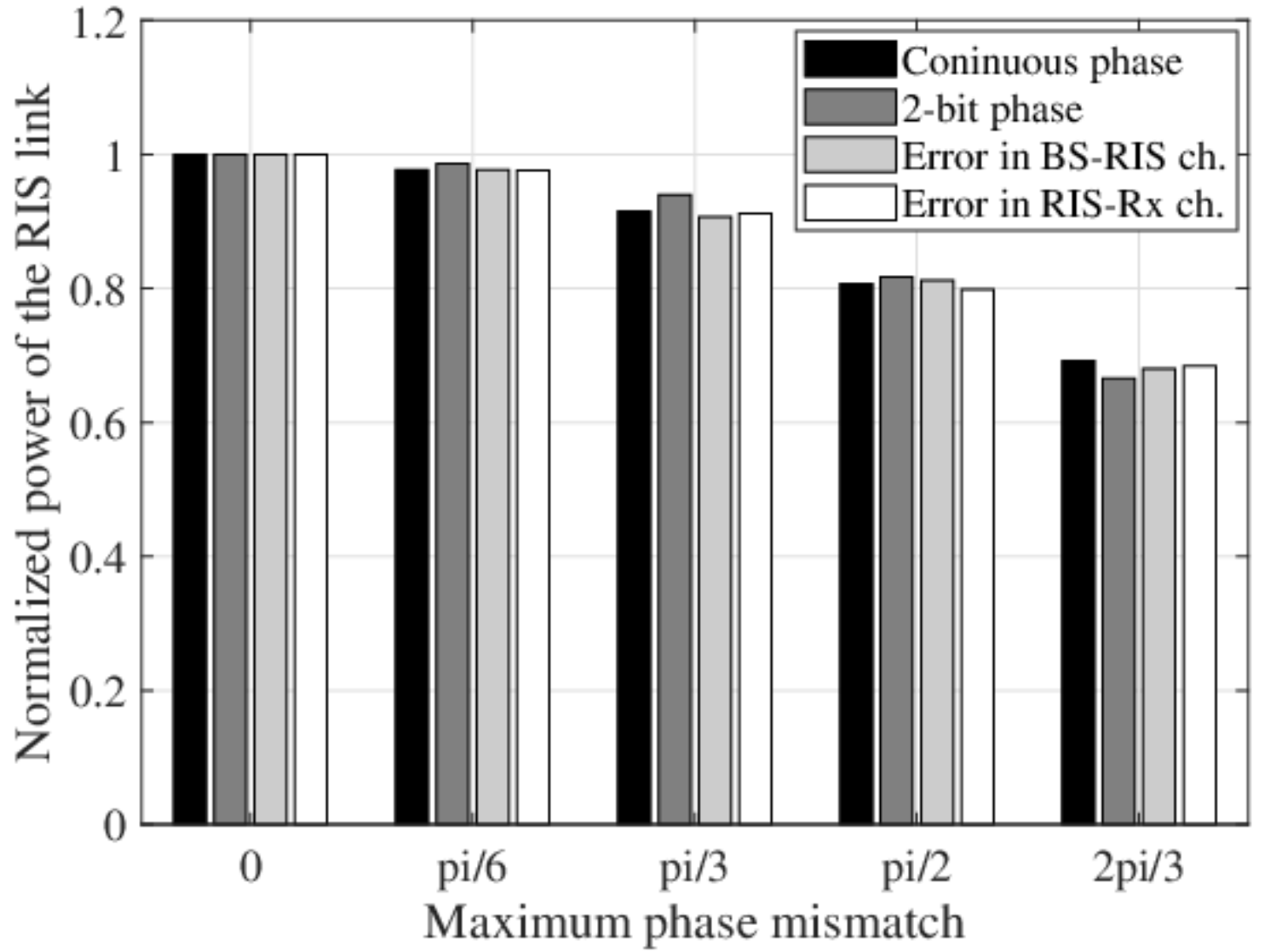}
    \caption{The normalized gain of the BS-RIS-Rx link for different values of maximum phase errors due to channel estimation error.}
    \label{fig:gainError}
\end{figure}

Fig.~\ref{fig:gainError} illustrates the power of the BS-RIS-Receiver link (i.e., $\Vert \mathbf{g}(t)^H \mathbf{\Theta}(t) \mathbf{h}(t)\Vert ^2$) when there is a mismatch in the phase shifts of the RIS elements, $\mathbf{\Theta}(t)$, due to channel estimation error. The power is normalized by the ideal case corresponding to perfect CSI estimation. The mismatch is modeled as a random variable uniformly distributed between $0$ and the maximum phase mismatch. The number of RIS elements, $N$ is $1024$. The first two bar plots respectively present the continuous phase and the case where the phase shifts of the RIS elements are assumed to take discrete values from a finite set, whose members are determined by the number of quantization bits. For instance, a two-bit quantization corresponds to four different phase shift values. In both cases, the error stems from estimating the cascaded effect of $\mathbf{g}$ and $\mathbf{h}$ collectively. The last two bar plots represent the cases where the estimation error only occurs at either of these two channels. It is observed from Fig.~\ref{fig:gainError} that the loss due to CSI estimation error is minimal for smaller estimation error values; it is less than $10\%$ for a phase mismatch of up to $\pi/3$. Furthermore, the impact of the CSI estimation error on the power gain of the RIS-assisted link will be further minimized when considered together with other non-idealities such as quantization error, absorption loss (i.e., $\beta_n < 1$), and mutual coupling between the RIS elements. This is evident from the plots corresponding to `2-bit phase', where we observe that some of the phase error due to channel estimation error is positively offset by the quantization error resulting in a slightly better performance compared to the `continuous phase' case. Although not reported here, a similar trend is observed for other values of $N$.

\subsection{Application of the \textit{ML and AI Algorithms} Block}
\label{sub:MLapplication}

In this last subsection, we extend the investigation scenario to multiple actuators and consider non-ideal RIS with absorption and mutual coupling loss and RIS phase shifts quantized into four levels, i.e., represented by $2-$bits. The amplitude of the RIS phase response is modeled following the empirical model presented in~\cite{AZW+20_RISamplitude}. The amplitude $\beta_n$, which is a function of $\theta_n$, is given as~\cite[(5)]{AZW+20_RISamplitude}
\begin{equation}
        \beta_{n}(\theta_{n}) = \left(1-\beta_{\min} \right) \left({\frac{\sin (\theta_{n} - \phi) +1}{2} }\right)^\alpha + \beta_{\min},
\end{equation}
where $\phi$ and $\alpha$ are constants related to the specific circuit implementation. We have set $\beta_{\min} = 0.8$. A multi-antenna BS concurrently serves four single-antenna actuators placed in an indoor factory area at the points $(135,105), (105,135), (120, 90)$ and $(90, 120)$, respectively. The BS is positioned at $(75, 75)$ and the RIS is located at the edge at $(150, 150)$. The (quantized) phase shifts of the RIS elements are chosen such that the sum rate at the four actuators is optimized, subject to a stringent outage probability constraint of $10^{-6}$ at each of the actuators. The ensuing optimization problem is not solvable using conventional methods due to the mutual interference coupling among the different actuators. Given the complexity of the optimization problem, we apply ML algorithms to solve it. In short, the Twin Delayed Deep Deterministic Policy Gradient (TD3) deep reinforcement learning algorithm~\cite{FHM18_TD3} is applied where the set of RIS phase shift values form the action space $\mathcal{A}$, and the sum rate considering the outage probability constraint is the reward function $\mathcal{R}$. The Shannon rate and the finite blocklength (FBL) rate~\cite{DKP16_shortPackets} for a blocklength of $20$ channel uses are considered. The latter is a more accurate representation of the achievable rate in factory automation use cases transmitting short but critical status update message. The application of \textit{ML and AI Algorithms} block to solve the complex problem of optimizing the RIS elements' phase shifts also absolves the BS from needing to estimate the CSI of the involved channels. 

\begin{figure}[ht]
    \centering
    \includegraphics[width=0.65\linewidth]{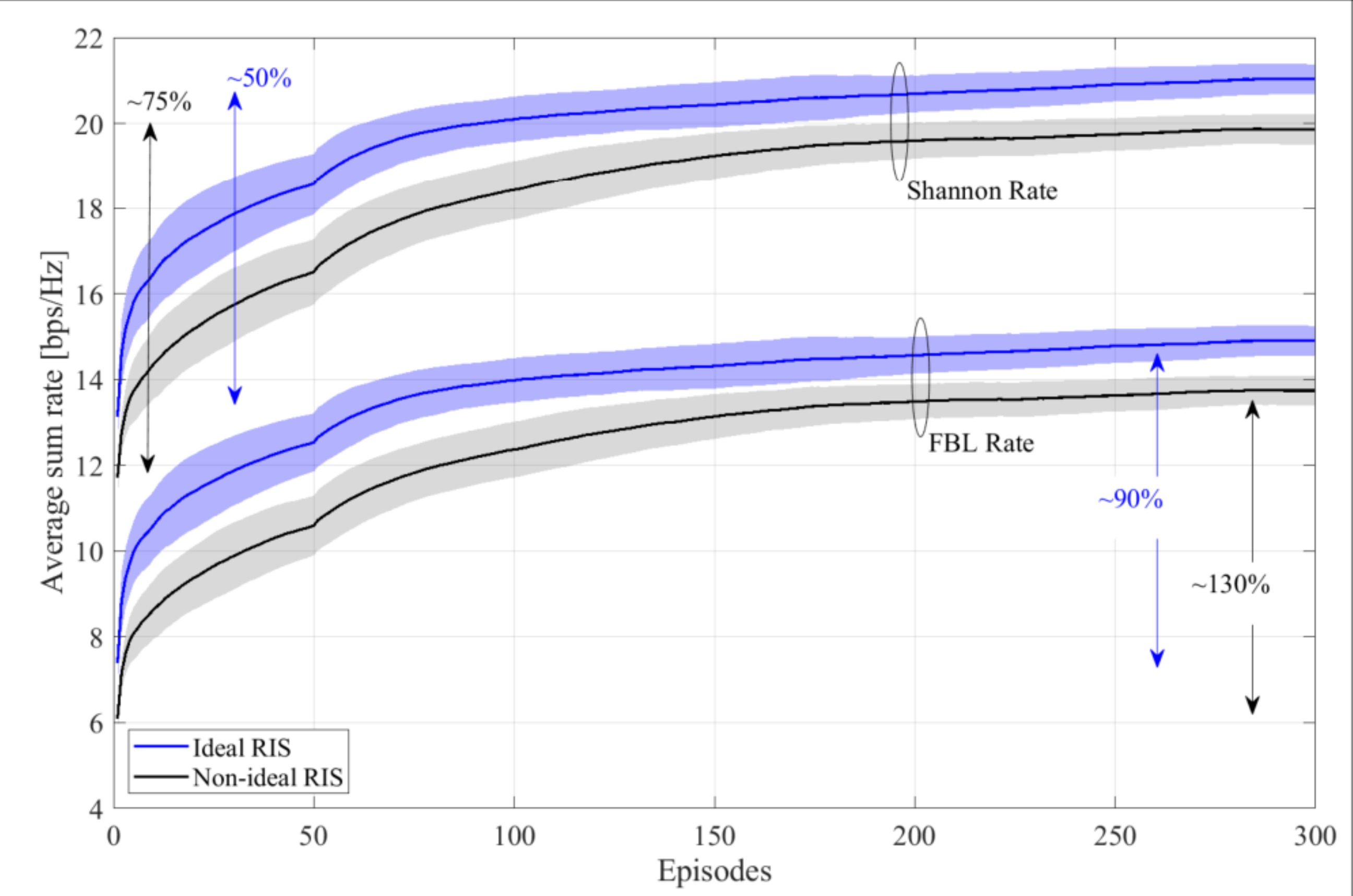}
    \caption{The average and the standard deviation of the sum Shannon and FBL rate of $4$ actuators in an RIS-assisted scenario with ideal and non-ideal RIS. The RIS phase shifts are optimized using the TD3 RL algorithm.}% Results are shown for an ideal, as well as a practical RIS element.} 
    \label{fig:rateWithTD3}
\end{figure}

Fig.~\ref{fig:rateWithTD3} presents the moving average (averaged over the past $50$ samples) of the sum rate at the four actuators as a function of the training episodes. The Shannon rate and the FBL rate considering an ideal RIS (i.e., continuous phase shifts) as well as non-ideal RIS with $\beta_{\min}=0.8$ and $2$-bit quantized phase shifts are shown by the blue and the black curves, respectively. The shadow represents the variation in the rate given by one standard deviation of the averaging samples. The limited number of samples to average over for Episodes $< 50$ results in the discontinuity observed around Episode $= 50$. We also see that there is a non-negligible loss in the sum rate of less than $10\%$ in the converged state resulting from the non-ideal assumptions, though the performance trend is the same. A significant gain in terms of the average sum rate is attained by optimizing the RIS elements phase shifts through the \textit{ML and AI algorithms} block. Approximate Shannon (FBL) rate gains of up to $50\% (90\%)$ and $75\% (130\%)$ are observed for ideal and non-ideal RIS elements, respectively. However, it must be noted that such gains are obtained when the ML algorithms converges over hundreds of episodes, which incurs non-negligible computational latency. Applications requiring low latency will impose constraints on the maximum number of episodes, and hence the obtained phase shift values may lead to more modest performance gains. 

\subsection{Summary}
The results presented in this section highlight the applicability of proposed system architecture in meeting the stringent QoS requirements of an IIoT network. We illustrated how an RIS can be employed to control the wireless environment and render the propagation condition more favorable. We have demonstrated how the \textit{Wireless Environment Control} block represented by an RIS would need to interact with other blocks, namely the \textit{Traffic/Channel Prediction} and the \textit{ML and AI Algorithms} blocks, to realize reliable and robust wireless links. Although not outlined here, the other blocks in the proposed functional architecture will be required to play a role in a more complex scenario. For instance, the \textit{Proactive Resource Management} and the \textit{E2E Optimization} blocks can introduce higher layer aspects which are not considered in the physical layer analysis presented here.

%%%%%%%%%%%%%%%%%%%%%%%%%%%%%%%
%%%%%%%%%%%%%%%%%%%%%%%%%%%%%%%
%%%                         %%%
%%%         SECTION         %%%
%%%                         %%%
%%%%%%%%%%%%%%%%%%%%%%%%%%%%%%%
%%%%%%%%%%%%%%%%%%%%%%%%%%%%%%%

\section{Conclusions}
\label{sec:conc}

This article proposes an innovative functional architecture for 6G special purpose IIoT networks comprising four distinct special purpose functionalities and three enabling technologies. The special-purpose functionalities are made up of RAN functions such as \textit{Wireless Environment Control} and \textit{Traffic/Channel Prediction}; and cross-layer optimized functionalities like \textit{Proactive Resource Management} and \textit{E2E Optimization}. The enabling technologies for these special-purpose functionalities include \textit{Synchronization and Coordination}, \textit{ML and AI Algorithms} and \textit{Auxiliary Functions}. The proposed architecture is designed to enable resource-efficient solutions to the complex and dynamically changing requirement of emerging IIoT applications. All these functionalities come at the cost of manageable communications overhead and a higher computational cost from the side of the network, which, thanks to cloud and edge computing technologies can be very efficiently managed. The applicability of the proposed architecture is demonstrated in a RIS-assisted wireless IIoT network where the phase shifts of the RIS elements residing in the \textit{Wireless Environment Control} block are optimized with the help of the \textit{Traffic/Channel Prediction} and ML and AI blocks to improve the sum rate performance under a stringent outage constraint. The numerical results demonstrate a sum throughput gain of over $100\%$ under realistic assumptions for the FBL rate. 


% % biography section
% % 
% % If you have an EPS/PDF photo (graphicx package needed) extra braces are
% % needed around the contents of the optional argument to biography to prevent
% % the LaTeX parser from getting confused when it sees the complicated
% % \includegraphics command within an optional argument. (You could create
% % your own custom macro containing the \includegraphics command to make things
% % simpler here.)
% %\begin{IEEEbiography}[{\includegraphics[width=1in,height=1.25in,clip,keepaspectratio]{mshell}}]{Michael Shell}
% % or if you just want to reserve a space for a photo:

% \end{IEEEbiography}


% % insert where needed to balance the two columns on the last page with
% %\newpage


% % You can push biographies down or up by placing
% % a \vfill before or after them. The appropriate
% % use of \vfill depends on what kind of text is
% % on the last page and whether or not the columns
% % are being equalized.

% %\vfill

% % Can be used to pull up biographies so that the bottom of the last one
% % is flush with the other column.
% %\enlargethispage{-5in}

\vspace{-10mm}
\begin{IEEEbiography}[{\includegraphics[width=1in,height=1.25in,clip,keepaspectratio]{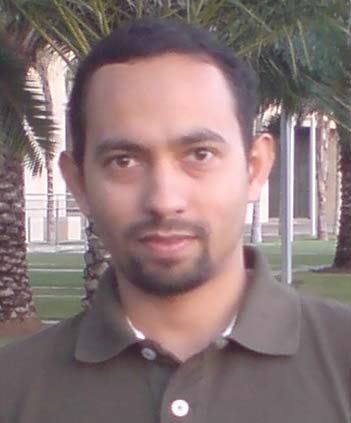}}]{Nurul Huda Mahmood} is a senior researcher and Adjunct Professor at Center for Wireless Communications (CWC), University of Oulu, where he is also the coordinator for Wireless Connectivity research area and leads critical MTC research within the 6G Flagship research program. His current research focus is on resilient communications for the 6G era.
\end{IEEEbiography}
\vspace{-12mm}
\begin{IEEEbiography}[{\includegraphics[width=1in,height=1.25in,clip,keepaspectratio]{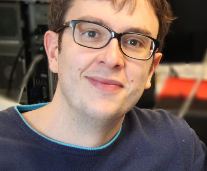}}]{Gilberto Berardinelli} (Senior Member, IEEE) received the first and second level degrees (cum laude) in telecommunication engineering from the University of L'Aquila, Italy, in 2003 and 2005, respectively, and the Ph.D. degree from Aalborg University, Denmark, in 2010. He is currently an Associate Professor with the Wireless Communication Networks (WCN) Section, Aalborg University, and also working in tight cooperation with Nokia Bell Labs. He is the author or coauthor of more than $100$ international publications, including conference proceedings, journal contributions, and book chapters. His research interests include physical layer, medium access control, and radio resource management design for 5G systems and beyond.
\end{IEEEbiography}

\begin{IEEEbiography}[{\includegraphics[width=1in,height=1.25in,clip,keepaspectratio]{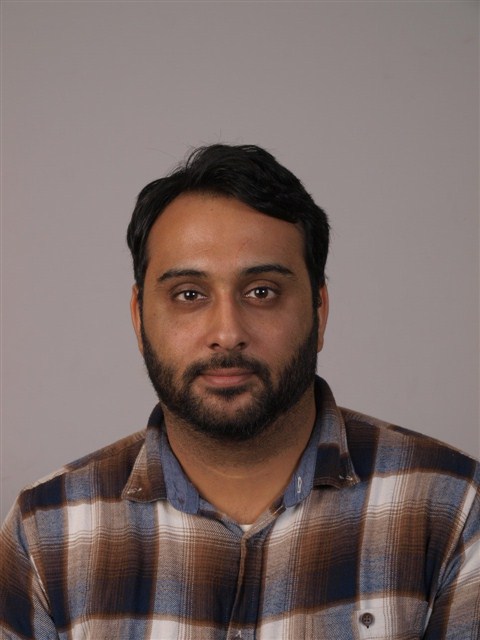}}]{Emil J. Khatib} is a postdoctoral researcher in the University of M\'{a}laga. He got a PhD in 2017 on the topic of Machine Learning, Big Data analytics and Knowledge Acquisition applied to the troubleshooting in cellular networks. He has participated in several national and international projects related to Industry 4.0 projects. Currently he is working on the topic of security, energy efficiency and localization in B5G/6G networks.
\end{IEEEbiography}
\vspace{-12mm}
\begin{IEEEbiography}[{\includegraphics[width=1in,height=1.25in,clip,keepaspectratio]{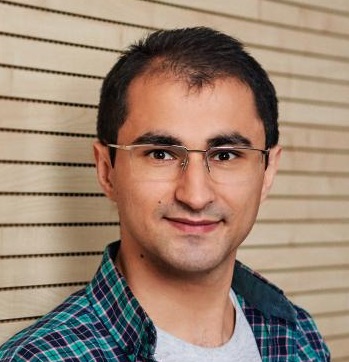}}]{Ramin Hashemi} (Student Member, IEEE) was born in Ardabil, Iran. He received the  B.Sc. and  M.Sc. degrees (with honors) from Amirkabir University of Technology (Tehran Polytechnic), Tehran, Iran, in 2016 and 2018, respectively. He is currently a doctoral student at the CWC, University of Oulu, Finland in the 6G Flagship project. His research interests include resource allocation in wireless networks, URLLC, RIS, MIMO wireless communications, beyond 5G, and fiber-optic communications.
\end{IEEEbiography}
\vspace{-12mm}
\begin{IEEEbiography}[{\includegraphics[width=1in,height=1.25in,clip,keepaspectratio]{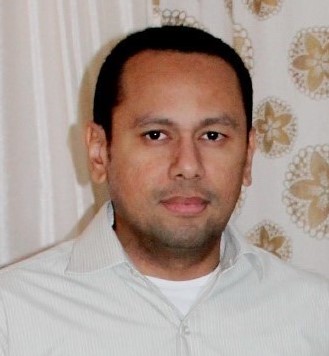}}]{Carlos De Lima} received the B.Sc. and M.Sc. in Electrical Engineering from the Federal University of Cear\'a (UFC), Brazil in 2002 and 2004, respectively. In 2013, he received the Dr.Sc (Tech) degree in communications engineering from the University of Oulu, Finland. From 2000 to 2005, he worked as research scientist in the Wireless Telecommunication Research Group (GTEL), Brazil. In 2005 he was a visiting researcher in the Ericsson Research Center, Lule\r{a}, Sweden. In 2006, he worked at the Nokia Institute of Technology (INdT), Brazil. From 2014 to 2018, he worked as an assistant professor at the Sao Paulo State University (UNESP), Brazil. From 2018 to 2021, he was with CWC, University of Oulu, Finland. He joined Nokia Mobile Networks in 2021 and currently investigates 5G systems and technologies.
\end{IEEEbiography}
\vspace{-12mm}
\begin{IEEEbiography}[{\includegraphics[width=1in,height=1.25in,clip,keepaspectratio]{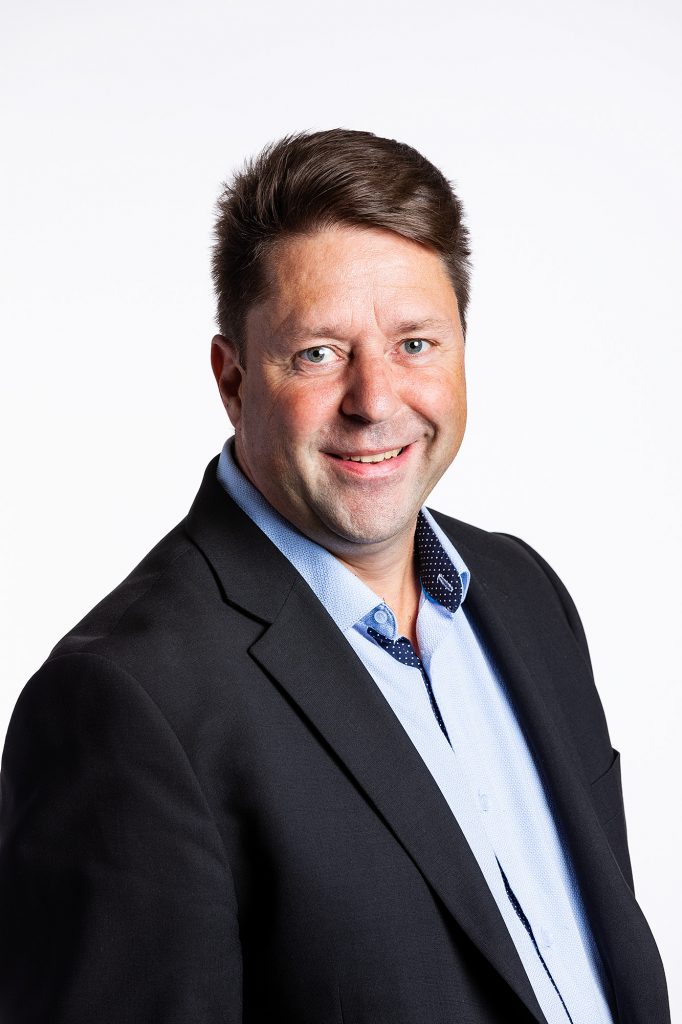}}]{Matti Latva-aho} received the M.Sc., Lic.Tech. and Dr. Tech (Hons.) degrees in Electrical Engineering from the University of Oulu, Finland in 1992, 1996 and 1998, respectively. From 1992 to 1993, he was a Research Engineer at Nokia Mobile Phones, Oulu, Finland after which he joined CWC at the University of Oulu. Prof. Latva-aho was Director of CWC during the years 1998-2006 and Head of Department for Communication Engineering until August 2014. Currently he is professor at the University of Oulu on wireless communications and Director for National 6G Flagship Programme. He is also a Global Research Fellow with Tokyo University. His research interests are related to mobile broadband communication systems and currently his group focuses on 6G systems research. Prof. Latva-aho has published over 500 conference or journal papers in the field of wireless communications. He received Nokia Foundation Award in 2015 for his achievements in mobile communications research.
\end{IEEEbiography}


\begin{thebibliography}{10}
\providecommand{\url}[1]{#1}
\csname url@samestyle\endcsname
\providecommand{\newblock}{\relax}
\providecommand{\bibinfo}[2]{#2}
\providecommand{\BIBentrySTDinterwordspacing}{\spaceskip=0pt\relax}
\providecommand{\BIBentryALTinterwordstretchfactor}{4}
\providecommand{\BIBentryALTinterwordspacing}{\spaceskip=\fontdimen2\font plus
\BIBentryALTinterwordstretchfactor\fontdimen3\font minus
  \fontdimen4\font\relax}
\providecommand{\BIBforeignlanguage}[2]{{%
\expandafter\ifx\csname l@#1\endcsname\relax
\typeout{** WARNING: IEEEtran.bst: No hyphenation pattern has been}%
\typeout{** loaded for the language `#1'. Using the pattern for}%
\typeout{** the default language instead.}%
\else
\language=\csname l@#1\endcsname
\fi
#2}}
\providecommand{\BIBdecl}{\relax}
\BIBdecl

\bibitem{MAS+22_factory5G}
A.~Mahmood \emph{et~al.}, ``Factory {5G}: A review of industry-centric features
  and deployment options,'' \emph{IEEE Industrial Electronics Magazine}, pp.
  2--12, Feb. 2022.

\bibitem{MTCwhitePaper2020}
N.~H. Mahmood \emph{et~al.}, \emph{White Paper on Critical and Massive Machine
  Type Communication towards {6G}}, ser. 6G Research Visions, nr. 11, N.~H.
  Mahmood \emph{et~al.}, Eds.\hskip 1em plus 0.5em minus 0.4em\relax Oulu,
  Finland: University of Oulu, Jun. 2020.

\bibitem{MBA+22_IIoT}
A.~Mahmood \emph{et~al.}, ``Industrial {IoT in 5G}-and-beyond networks: Vision,
  architecture, and design trends,'' \emph{IEEE Transactions on Industrial
  Informatics}, vol.~18, no.~6, pp. 4122--4137, Jun. 2022.

\bibitem{MMG+21_IIoT}
N.~H. Mahmood, N.~Marchenko, M.~Gidlund, and P.~Popovski, Eds., \emph{Wireless
  Networks and Industrial IoT: Applications, Challenges and Enablers}.\hskip
  1em plus 0.5em minus 0.4em\relax Zurich, Switzerland: Springer, 2021.

\bibitem{berardinelli2018_wirt}
G.~Berardinelli, N.~H. Mahmood, I.~Rodriguez, and P.~E. Mogensen, ``Beyond {5G}
  wireless {IRT} for {Industry} 4.0: Design principles and spectrum aspects,''
  in \emph{Proc. IEEE Globecom Workshops}, Abu Dhabi, UAE, Dec. 2018.

\bibitem{CLH+21_multirobot}
K.-C. Chen \emph{et~al.}, ``Wireless networked multirobot systems in smart
  factories,'' \emph{Proceedings of the IEEE}, vol. 109, no.~4, pp. 468--494,
  Apr. 2021.

\bibitem{adeogun2020towards}
R.~Adeogun \emph{et~al.}, ``Towards {6G in-X} subnetworks with sub-millisecond
  communication cycles and extreme reliability,'' \emph{IEEE Access}, vol.~8,
  pp. 110\,172--110\,188, 2020.

\bibitem{6gWP2019}
M.~Latva-aho and K.~Lepp\"{a}nen~(ed.), \emph{Key drivers and research
  challenges for {6G} ubiquitous wireless intelligence (white paper)}.\hskip
  1em plus 0.5em minus 0.4em\relax Oulu, Finland: 6G Flagship, Sep. 2019.

\bibitem{DAS_19_whatShould6G}
S.~Dang, O.~Amin, B.~Shihada, and M.-S. Alouini, ``What should {6G} be?''
  \emph{Nature Electronics}, vol.~3, no.~1, pp. 20--29, Jan. 2020.

\bibitem{MBM+21_mtcEurasip}
N.~H. Mahmood \emph{et~al.}, ``Machine type communications: key drivers and
  enablers towards the {6G} era,'' \emph{EURASIP Journal on Wireless
  Communications and Networking}, vol. 2021, no. 134, pp. 1--25, Jun. 2021.

\bibitem{URB+21_hexaX_6G}
M.~A. Uusitalo \emph{et~al.}, ``{6G} vision, value, use cases and technologies
  from european {6G} flagship project {Hexa-X},'' \emph{IEEE Access}, vol.~9,
  pp. 160\,004--160\,020, Nov. 2021.

\bibitem{viswanath6G_2020}
H.~{Viswanathan} and P.~E. {Mogensen}, ``Communications in the {6G} era,''
  \emph{IEEE Access}, vol.~8, pp. 57\,063--57\,074, Mar. 2020.

\bibitem{samsung6g}
\BIBentryALTinterwordspacing
{Samsung Research}, ``The next hyper-connected experience for all,'' Jul. 2020,
  white paper. [Online]. Available:
  \url{https://research.samsung.com/next-generation-communications}
\BIBentrySTDinterwordspacing

\bibitem{GMB19:5Gevolution}
A.~{Ghosh}, A.~{Maeder}, M.~{Baker}, and D.~{Chandramouli}, ``{5G} evolution: A
  view on {5G} cellular technology beyond {3GPP} release 15,'' \emph{IEEE
  Access}, vol.~7, pp. 127\,639--127\,651, 2019.

\bibitem{LCM+22_ISAC}
F.~Liu \emph{et~al.}, ``Integrated sensing and communications: Towards
  dual-functional wireless networks for {6G} and beyond,'' \emph{IEEE Journal
  on Selected Areas in Communications}, Mar. 2022.

\bibitem{khatib2021optimization}
E.~J. Khatib and R.~Barco, ``Optimization of {5G} networks for smart
  logistics,'' \emph{Energies}, vol.~14, no.~6, p. 1758, 2021.

\bibitem{ahokangas2021platform}
P.~Ahokangas, M.~Matinmikko-Blue, S.~Yrj{\"o}l{\"a}, and H.~H{\"a}mm{\"a}inen,
  ``Platform configurations for local and private {5G} networks in complex
  industrial multi-stakeholder ecosystems,'' \emph{Telecommunications Policy},
  vol.~45, no.~5, p. 102128, 2021.

\bibitem{NHS+21_dependability}
N.~Franchi \emph{et~al.}, ``Selected aspects and approaches on improving
  dependability in industrial radio networks,'' in \emph{Wireless Networks and
  Industrial IoT: Applications, Challenges and Enablers}, N.~H. Mahmood,
  N.~Marchenko, M.~Gidlund, and P.~Popovski, Eds.\hskip 1em plus 0.5em minus
  0.4em\relax Zurich, Switzerland: Springer, 2021, pp. 21--38.

\bibitem{MAM+22_missionEffCap}
I.~Muhammad \emph{et~al.}, ``Mission effective capacity—a novel dependability
  metric: A study case of multiconnectivity-enabled {URLLC for IIoT},''
  \emph{IEEE Transactions on Industrial Informatics}, vol.~18, no.~6, pp.
  4180--4188, Jun. 2022.

\bibitem{GRN21_tsn}
D.~Ginth{\"o}r, Ren{\'e}, N.~Nayak, and J.~von Hoyningen-Huene,
  ``Time-sensitive networking for industrial control networks,'' in
  \emph{Wireless Networks and Industrial IoT: Applications, Challenges and
  Enablers}, N.~H. Mahmood, N.~Marchenko, M.~Gidlund, and P.~Popovski,
  Eds.\hskip 1em plus 0.5em minus 0.4em\relax Zurich, Switzerland: Springer,
  2021, pp. 39--54.

\bibitem{hexaXd1_3}
{Hexa-X}, ``Targets and requirements for {6G} - initial {E2E} architecture,''
  Feb. 2022, deliverable D1.3.

\bibitem{5GACIA_integration5G_tsn}
5G-ACIA, ``Integration of {5G} with time-sensitive networking for industrial
  communications,'' Feb. 2021.

\bibitem{mahmood2019resource}
N.~H. Mahmood \emph{et~al.}, ``On the resource utilization of
  multi-connectivity transmission for urllc services in {5G New Radio},'' in
  \emph{2019 IEEE Wireless Communications and Networking Conference Workshop
  (WCNCW)}.\hskip 1em plus 0.5em minus 0.4em\relax IEEE, 2019, pp. 1--6.

\bibitem{ZX21_channelKnowledgeMapping}
Y.~Zeng and X.~Xu, ``Toward environment-aware {6G} communications via channel
  knowledge map,'' \emph{IEEE Wireless Communications}, vol.~28, no.~3, pp.
  84--91, Jun. 2021.

\bibitem{DZD+20_RIS}
M.~{Di Renzo} \emph{et~al.}, ``Smart radio environments empowered by
  reconfigurable intelligent surfaces: How it works, state of research, and the
  road ahead,'' \emph{IEEE Journal on Selected Areas in Communications},
  vol.~38, no.~11, pp. 2450--2525, Nov. 2020.

\bibitem{MLA+20_predictive}
N.~H. {Mahmood}, O.~A. {Lopez}, H.~{Alves}, and M.~{Latva-aho}, ``A predictive
  interference management algorithm for {URLLC} in beyond {5G} networks,''
  \emph{IEEE Communications Letters}, vol.~25, no.~3, pp. 995--999, Mar. 2021.

\bibitem{wang2015approach}
S.~Wang \emph{et~al.}, ``An approach for spatial-temporal traffic modeling in
  mobile cellular networks,'' in \emph{2015 27th International Teletraffic
  Congress}, 2015, pp. 203--209.

\bibitem{HM20_lstm}
M.~S. {Hossain} and G.~{Muhammad}, ``A deep-tree-model-based radio resource
  distribution for {5G} networks,'' \emph{IEEE Wireless Communications},
  vol.~27, no.~1, pp. 62--67, Feb. 2020.

\bibitem{ZZY+18_cnn}
C.~{Zhang}, H.~{Zhang}, D.~{Yuan}, and M.~{Zhang}, ``Citywide cellular traffic
  prediction based on densely connected convolutional neural networks,''
  \emph{IEEE Communications Letters}, vol.~22, no.~8, pp. 1656--1659, Aug.
  2018.

\bibitem{BGF+22_networkIntelligence}
A.~Banchs, A.~Garcia-Saavedra, M.~Fiore, and M.~Gramaglia, ``Network
  intelligence in {6G}: challenges and opportunities,'' in \emph{inProc. 16th
  Workshop on Mobility in the Evolving Internet Architecture (MobiArch) 2021},
  New Orleans, USA, Jan. 2022, pp. 7--12.

\bibitem{BVS+20_slicingRA}
A.~{Banchs}, G.~{de Veciana}, V.~{Sciancalepore}, and X.~{Costa-Perez},
  ``Resource allocation for network slicing in mobile networks,'' \emph{IEEE
  Access}, vol.~8, pp. 214\,696--214\,706, Nov. 2020.

\bibitem{FP20_NSmeetsPT}
R.~{Fantacci} and B.~{Picano}, ``When network slicing meets prospect theory: A
  service provider revenue maximization framework,'' \emph{IEEE Transactions on
  Vehicular Technology}, vol.~69, no.~3, pp. 3179--3189, Mar. 2020.

\bibitem{KFV+21_E2Eco}
J.~Kotary, F.~Fioretto, P.~Van~Hentenryck, and B.~Wilder, ``End-to-end
  constrained optimization learning: A survey,'' in \emph{Proceedings of the
  Thirtieth International Joint Conference on Artificial Intelligence
  (IJCAI'21)}, Montreal, Canada, Aug. 2021, p. 4475–4482.

\bibitem{abbas2017mobile}
N.~Abbas, Y.~Zhang, A.~Taherkordi, and T.~Skeie, ``Mobile edge computing: A
  survey,'' \emph{IEEE Internet of Things Journal}, vol.~5, no.~1, pp.
  450--465, 2017.

\bibitem{art:godor-icsm20}
I.~{Godor} \emph{et~al.}, ``A look inside {5G} standards to support time
  synchronization for smart manufacturing,'' \emph{IEEE Communications
  Standards Magazine}, vol.~4, no.~3, pp. 14--21, 2020.

\bibitem{khatib2016self}
E.~J. Khatib \emph{et~al.}, ``Self-healing in mobile networks with big data,''
  \emph{IEEE Communications Magazine}, vol.~54, no.~1, pp. 114--120, 2016.

\bibitem{rai2020explainable}
A.~Rai, ``Explainable ai: From black box to glass box,'' \emph{Journal of the
  Academy of Marketing Science}, vol.~48, no.~1, pp. 137--141, 2020.

\bibitem{mika_wp10}
\BIBentryALTinterwordspacing
M.~Ylianttila \emph{et~al.}, Eds., \emph{{6G} White Paper: Research Challenges
  for Trust, Security and Privacy}, ser. 6G Research Visions, nr. 9.\hskip 1em
  plus 0.5em minus 0.4em\relax Oulu, Finland: University of Oulu, Jun. 2020.
  [Online]. Available: \url{http://jultika.oulu.fi/files/isbn9789526226804.pdf}
\BIBentrySTDinterwordspacing

\bibitem{HAM+22_jointRIS_ICL}
R.~Hashemi, S.~Ali, N.~H. Mahmood, and M.~Latva-aho, ``Joint sum rate and
  blocklength optimization in {RIS}-aided short packet {URLLC} systems,''
  \emph{IEEE Communications Letters}, 2022.

\bibitem{Che16_industrialChannel}
M.~Cheffena, ``Propagation channel characteristics of industrial wireless
  sensor networks [wireless corner],'' \emph{IEEE Antennas and Propagation
  Magazine}, vol.~58, no.~1, pp. 66--73, Feb. 2016.

\bibitem{BDD+19_RIS}
E.~Basar \emph{et~al.}, ``Wireless communications through reconfigurable
  intelligent surfaces,'' \emph{IEEE Access}, vol.~7, pp. 116\,753--116\,773,
  Aug. 2019.

\bibitem{WSD21_risChannel}
X.~Wei, D.~Shen, and L.~Dai, ``Channel estimation for {RIS} assisted wireless
  communications—part {I}: Fundamentals, solutions, and future
  opportunities,'' \emph{IEEE Communications Letters}, vol.~25, no.~5, pp.
  1398--1402, May 2021.

\bibitem{MA21_chEstRIS}
J.~Mirza and B.~Ali, ``Channel estimation method and phase shift design for
  reconfigurable intelligent surface assisted {MIMO} networks,'' \emph{IEEE
  Transactions on Cognitive Communications and Networking}, vol.~7, no.~2, pp.
  441--451, Feb. 2021.

\bibitem{HAM+21_RIS_URLLC}
R.~Hashemi, S.~Ali, N.~H. Mahmood, and M.~Latva-aho, ``Average rate and error
  probability analysis in short packet communications over {RIS}-aided {URLLC}
  systems,'' \emph{IEEE Transactions on Vehicular Technology}, vol.~70, no.~10,
  pp. 10\,320--10\,334, Oct. 2021.

\bibitem{AZW+20_RISamplitude}
S.~Abeywickrama, R.~Zhang, Q.~Wu, and C.~Yuen, ``Intelligent reflecting
  surface: Practical phase shift model and beamforming optimization,''
  \emph{IEEE Transactions on Communications}, vol.~68, no.~9, pp. 5849--5863,
  Sep. 2020.

\bibitem{FHM18_TD3}
S.~Fujimoto, H.~van Hoof, and D.~Meger, ``Addressing function approximation
  error in actor-critic methods,'' in \emph{inProc. 35th International
  Conference on Machine Learning}, Stockholm, Sweden, Jul. 2018, pp.
  1587--1596.

\bibitem{DKP16_shortPackets}
G.~Durisi, T.~Koch, and P.~Popovski, ``Toward massive, ultrareliable, and
  low-latency wireless communication with short packets,'' \emph{Proceedings of
  the IEEE}, vol. 104, no.~9, pp. 1711--1726, Sep. 2016.

\end{thebibliography}
\end{document}